\documentclass[aps,prx,twocolumn,amsmath,amssymb,showpacs,superscriptaddress,notitlepage]{revtex4-2}
\usepackage{amsmath,amssymb,amsfonts,bm}
\usepackage{graphicx}
\usepackage{dcolumn}
\usepackage{mathrsfs}
\usepackage{bbold}
\usepackage{dsfont}
\usepackage{dcolumn}
\usepackage{epstopdf}
\usepackage[colorlinks=true,linkcolor=blue,citecolor=blue, urlcolor=blue,bookmarks=false]{hyperref}
\usepackage{changes}
\usepackage{textgreek}

\begin{document}

\title{Altermagnetism in quasicrystals}

\author{Rui Chen}
\affiliation{Department of Physics, Hubei University, Wuhan 430062, China}

\author{Bin Zhou}
\affiliation{Department of Physics, Hubei University, Wuhan 430062, China}

\author{Dong-Hui Xu}\email[]{donghuixu@cqu.edu.cn}
\affiliation{Department of Physics and Chongqing Key Laboratory for Strongly Coupled Physics, Chongqing University, Chongqing 400044, China}
\affiliation{Center of Quantum Materials and Devices, Chongqing University, Chongqing 400044, China}

\begin{abstract}
Altermagnets are a recently discovered class of magnetic materials that combine a collinear, zero-magnetization spin structure, characteristic of antiferromagnets, with spin-split electronic bands, a hallmark of ferromagnets. This unique behavior arises from the breaking of combined time-reversal and spatial symmetries (such as inversion or lattice translation), which are preserved in conventional antiferromagnets. To date, research has mainly focused on altermagnetic phases in periodic crystals, where the order is linked to rotational symmetries compatible with translational periodicity. In this Letter, we demonstrate that quasicrystals, which possess rotational symmetries incompatible with periodicity, can host exotic altermagnetic orders. Using symmetry analysis and self-consistent mean-field theory, we predict stable $g$-wave and $i$-wave altermagnetism in octagonal and dodecagonal quasicrystals, respectively. These phases are characterized by global $C_8 T$ and $C_{12} T$ symmetries and exhibit anisotropic spin-splittings in their spectral functions and spin conductance, with characteristic eight- and twelve-fold nodal structures that establish a theoretical framework for identifying these phases in future experiments. Our findings establish quasicrystals as a versatile platform for realizing unconventional altermagnetic orders beyond the constraints of periodicity.
\end{abstract}
\maketitle

{\emph{Introduction}.}---Altermagnets represent a recently discovered class of magnetic materials, distinct from conventional ferromagnets and antiferromagnets~\cite{Smejkal22PRX,Bai2024AFM,Naka19NC,Ahn19PRB,
hayami2019momentum,Yuan2020Giant,mazin2021prediction,
ma2021multifunctional,
ifmmode2022Emerging,krempasky2024altermagnetic,mejkal2022NatRevMat}. They uniquely combine a collinear spin structure with zero net magnetization, characteristic of antiferromagnets, with spin-split electronic bands, a hallmark of ferromagnets. This unconventional behavior is rooted in the breaking of composite symmetries—specifically, time-reversal combined with spatial operations like inversion or translation —which remain intact in conventional antiferromagnets. It enables unique physical phenomena such as anomalous transport~\cite{Nakaprb,shao2021spin,Fernandestoplogical,zhang2024prl,Zhou2024PRL}, giant tunneling magnetoresistance~\cite{sdongFeSb2,ifmmode2022Giant}, and potential topological responses~\cite{Ezawa2024PRLAlter,LiuCC2024PRB,ChenR2024altermagnet,ChenR2025altermagnet}, thereby prompting widespread research both theoretically and experimentally~\cite{Bai2023PRL,Gonzalez2021PRL,vsmejkal2020crystal,Feng2022NatElec, Fedchenko2024SciAdv,Karube2022PRL,Keler2024NpjS}. To date, however, the exploration of altermagnetism has been almost exclusively confined to periodic crystals. Within this paradigm, altermagnetic phases are protected by specific crystal rotational symmetries~(e.g., four-fold or six-fold). This reliance on crystallographic symmetry fundamentally forbids altermagnets with higher-order or rotational symmetries incompatible with periodicity. This limitation necessitates the search for alternative platforms beyond periodic crystals to realize more exotic altermagnetic states.

\begin{figure}[t!]
	\centering
	\includegraphics[width=\columnwidth]{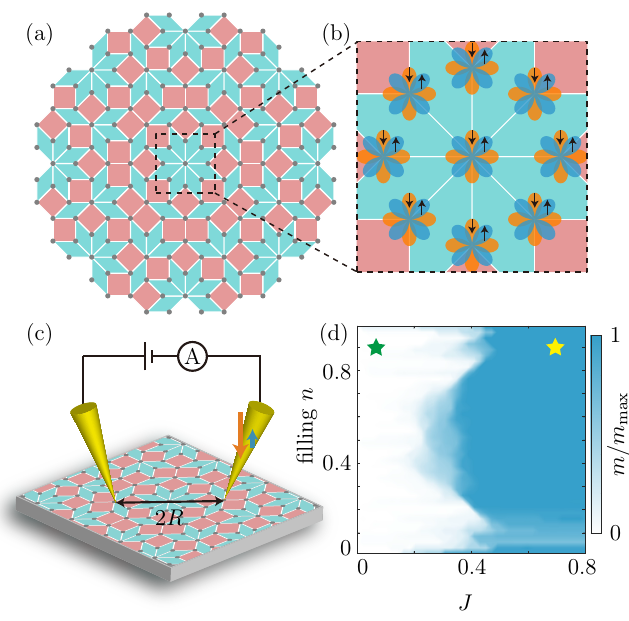}
	\caption{(a) Schematic illustration of the AB-tiling octagonal quasicrystal, which consists of two types of primitive tiles: square tiles (red) and rhombus tiles (cyan) with an acute angle $45^{\circ}$. (b) Each site of the AB-tiling quasicrystal features two orbitals (blue and orange) connected by a $C_8$ rotation. (c) We propose that the altermagnetic order in a quasicrystal can be probed by a double-tip STM setup. When a spin-unpolarized current is injected, the output current becomes spin-polarized due to the altermagnetic structure. (d) The normalized mean altermagnetization $m/m_{\max}$ as a function of the filling $n$ and the interaction coupling strength $J$. Each data point is obtained by using the self-consistent mean-field method. }
\label{fig_AB_phase}
\vspace{0.3 cm}
\end{figure}

Concurrently, a second, independent paradigm shift has transformed our understanding of aperiodic systems. Quasicrystals were considered textbook examples of geometrically frustrated structures~\cite{Grunbaum1987book,Steurer2007QC}, presumed to support only disordered, spin-glass-like magnetic states~\cite{Berger1990PRL,Goldman1991ARPC,Goldman2013NatMat}.However, subsequent theoretical studies ruled out the
notion that geometrical frustration fundamentally forbids long-range magnetic order in
quasicrystals~\cite{spingroupinQCs,Lifshitz2024ECMP,Lifshitz1997RMP,Lifshitz2000MSEA}.  Experimental realization of such phases remained elusive for
decades. The subsequent observation of long-range ferromagnetic order~\cite{Takeuchi2023PRL,Tamura2021JACS} and, most strikingly, long-range antiferromagnetic order in a true icosahedral quasicrystal~\cite{Tamura2025NatPhys,Labib2025PRB} has demonstrated that the complex aperiodic order of quasicrystals
can support coherent magnetic ground states. These breakthroughs position quasicrystals
as a fertile platform for discovering unconventional magnetism.

This work unites these two revolutionary frontiers by showing that the rotational symmetries of quasicrystals, which are incompatible with periodicity, can enable and stabilize forms of altermagnetic order unavailable in periodic crystals. This mechanism is distinct from recent proposals of altermagnetism in amorphous solids~\cite{Dornellas2025arXiv}, where rotational invariance is statistical rather than an exact long-range symmetry of a deterministic quasiperiodic structure. In the quasicrystalline case studied here, the altermagnetic order is protected by magnetic rotational symmetries tied to the quasiperiodic long-range order itself.

 Here we show that this principle enables altermagnetic phases with symmetry structures inaccessible in periodic crystals. Specifically, the Ammann-Beenker~(AB) tiling, a canonical octagonal quasicrystal, supports a $g$-wave altermagnetism protected by a composite eight-fold rotational symmetry $C_8T$, where $T$ is time-reversal symmetry. Our analysis of the Fourier spectra reveals a momentum-dependent spin structure with a nodal $g$-wave configuration. We propose that this unique order can be detected via double-tip scanning tunneling microscopy (STM)~\cite{Leeuwenhoek2020Nanoeng,Matsui2007RSI,Nakanishi2010PE,Hus2017PRL,Su2025arXiv} and show that transport measurements would exhibit a spin-polarized current whose anisotropy directly reflects the underlying $g$-wave symmetry. Furthermore, we extend our analysis to show that an $i$-wave altermagnetic phase can emerge in the Stampfli-tiling dodecagonal quasicrystal. Our findings establish a previously uncharted direction in magnetism, demonstrating that the rich symmetries of quasicrystals offer a new route to designing and discovering altermagnetic orders beyond periodic systems.

{\emph{Model}.}---We adopt a minimal $t$-$J$-like model~\cite{Dornellas2025arXiv} for the orbital altermagnet phase of an AB-tiling quasicrystal~[Fig.~\ref{fig_AB_phase}(a)]. The AB-tiling quasicrystal consists of two types of primitive tiles: square tiles (red) and rhombus tiles (cyan) with an acute angle of $45^{\circ}$. Each site contains two orbitals connected by a $C_8$ rotation~[Fig.~\ref{fig_AB_phase}(b)].  A concrete realization of this setup is achieved with the $\{d_{xy}, d_{x^2-y^2}\}$ orbitals. They form a symmetry-protected degenerate doublet belonging to the $E_2$ irreducible representation in the $C_{8v}$ point symmetry of the Ammann-Beenker tiling~\cite{Socolar1989PRB,Madison2013PSS}. The total Hamiltonian, $H=H_K+H_{\text{Int}}$, consists of a kinetic term and an interaction term. The kinetic term
\begin{equation}
H_K=\sum_{\langle j k\rangle} \Psi_j^{\dagger}\left[F\left(\eta\theta_{j k}\right) \right]\sigma_0 \Psi_k,
\label{Eq:Hamiltonian0}
\end{equation}
describes hopping between adjacent sites. Here, $\langle\cdots\rangle$ denotes sites connected by the bonds of distance $a=1$, as shown in Fig.~\ref{fig_AB_phase}(a). $\Psi_j^{\dagger}=\left(c_{j \alpha \uparrow}^{\dagger}, c_{j \beta \uparrow}^{\dagger}, c_{j \alpha \downarrow}^{\dagger}, c_{j \beta \downarrow}^{\dagger}\right)$ are electron creation
operators at site $j$.  $\alpha$ and $\beta$ represent the two orbitals at one site and spin degrees of freedom are denoted as $\uparrow$ and $\downarrow$.
$\theta_{j k}$ is the
polar angle of bond connecting sites $j$ and $k$ with respect to the horizontal direction. The orbital-dependent hopping matrix is given by
\begin{equation}
F(\eta\theta)=\left(\begin{array}{cc}
\left(t_1-t_2\right) \cos ^2 \eta\theta+t_2 & \left(t_1-t_2\right) \sin \eta\theta \cos \eta\theta \\
\left(t_1-t_2\right) \sin \eta\theta \cos \eta\theta & \left(t_1-t_2\right) \sin ^2 \eta\theta+t_2
\end{array}\right).
\label{Eq:hopping}
\end{equation}
We fix $t_1=1$, $t_2=1 / 2$, and $\eta=2$ in the case of the AB-tiling quasicrystal. At $\theta=0$, the hopping is dominated by the overlap between the first orbitals $t_1$, while the second orbitals have a smaller overlap $t_2$, with $t_2<t_1$. At $\theta=\pi/4$, these hopping strengths are reversed. This angular dependence captures the rotating orbital texture illustrated in Fig.~\ref{fig_AB_phase}(b), where the orbital character naturally twists along different bond directions due to the underlying $C_8$ rotational symmetry.

We include a ferromagnetic Heisenberg-like combined spin-orbital interaction~\cite{Giuli2025PRB,Dornellas2025arXiv}
\begin{equation}
H_{\text {Int }}=-J \sum_{\substack{\langle j k\rangle ,\zeta \chi}}\left(\Psi_j^{\dagger} \tau^\zeta  \sigma^\chi \Psi_j\right)\left(\Psi_k^{\dagger} \tau^\zeta  \sigma^\chi \Psi_k\right)-n_j n_k,
\label{Eq:interaction}
\end{equation}
where $n_j=\sum_{\mu,s}c_{j \mu s}^{\dagger} c_{j \mu s}$ is the total on-site occupation operator, with $s=\uparrow,\downarrow$ labels spin and $\mu=\alpha,\beta$ labels orbital. $\sigma^\chi\left(\tau^\zeta\right)$ are Pauli matrices acting on the spin (orbital) subspace. In the single-fermion-per-site limit, Eq.~\eqref{Eq:interaction} resembles a ferromagnetic Kugel-Khomskii interaction, expressed as $\left( \bm{S} _j \cdot \bm{S} _k\right)\left( \bm{\tau} _j \cdot \bm{\tau} _k\right)$~\cite{Kugel1982SPU}.  In Sec. SI of the Supplemental Material~\cite{Supp}, we provide a detailed group-theoretical derivation of the orbital selection and Slater-Koster hopping textures and discuss possible experimental directions.  Mn-based octagonal quasicrystals and approximants~\cite{Jiang1991PRL} provide useful motivation for octagonal quasiperiodic environments, although the Ammann-Beenker model used here should be viewed as a minimal symmetry-based model rather than a material-specific description of Mn-Si-Al.

{\emph{Mean-Field Decoupling and emergent altermagnetism}.}---The interaction term is decoupled via a mean-field approximation, yielding an effective single-particle Hamiltonian
\begin{equation}
H_{\text {Int }}=-J \sum_{\langle j k\rangle}\left\langle m_j\right\rangle m_k-\left\langle n_j\right\rangle n_k+(j \leftrightarrow k),
\label{Eq:interaction1}
\end{equation}
where $n_j$ is the total occupation and $m_j=\Psi_j^{\dagger} \sigma^z \tau^z \Psi_j$ denotes the local magnetization.  We assume spontaneous symmetry breaking occurs along  the $\sigma^z$ and orbital $\tau^z$ directions, retaining only the diagonal components relevant to the symmetry-broken state. At each iteration, we compute site-resolved expectation values for the local charge density $\langle n_j \rangle$ and magnetization $\langle m_j \rangle$. These expectation values are then used to update the mean-field Hamiltonian for the next iteration. This self-consistent procedure is repeated until convergence is achieved (typically within 200 iterations) and allows for spatially inhomogeneous ordering without presupposing translational symmetry.

This interaction term in Eq.~\eqref{Eq:interaction1} lifts the degeneracy among the four spin-orbital configurations. For a positive $m_j$, the states $|\alpha \uparrow \rangle$ and $|\beta \downarrow \rangle$ on the $j$-th site have lower energy, whereas $|\beta \uparrow \rangle$ and $|\alpha \downarrow \rangle$ have higher energy. Consequently, the spin and orbital degrees of freedom become locked, with electrons in the $\alpha$-orbital predominantly carrying spin-up and those in the $\beta$-orbital carrying spin-down [see Fig.~\ref{fig_AB_phase}(b)]. This establishes the altermagnetic phase in quasicrystals. Although the magnetic term $m_j$ individually breaks the eightfold rotation symmetry $C_8=\tau_y\sigma_0 \mathcal{R}_8$ and time-reversal symmetry $T=i\tau_0\sigma_y\mathcal{K}$, their combination $C_8 T$ remains a preserved symmetry of the system. Here, $\mathcal{R}_8$ is the orthogonal matrix that permutes the tiling sites to rotate the whole system by $\pi/4$ and $\mathcal{K}$ is complex conjugation. This composite symmetry plays a central role in characterizing the altermagnetic phase.

We emphasize that the full Hamiltonian lacks global spin-rotational symmetry due to the locking of the spin degrees of freedom to the fixed, geometry-dependent orbital textures of the quasicrystalline lattice. Consequently, the total energy of the system is sensitive to the specific direction of the spin and orbital polarization. In Sec. SIV of the Supplemental Material~\cite{Supp}, we performed fully unrestricted self-consistent calculations, and found that the altermagnetic state further breaks spin-rotational symmetry, with spin polarization stabilized along the principal axes ($x, y,$ or $z$) concomitant with orbital polarization ($\tau_z$). This specific alignment allows the system to preserve the combined $C_8 \mathcal{T}$ symmetry, where the spatial rotation $C_8$ acts on the lattice and the time-reversal operation $\mathcal{T}$ acts on the spins, which protects the characteristic $g$-wave spectral splitting.

{\emph{Interaction-induced altermagnetic ordering}.}---Figure~\ref{fig_AB_phase}(d) presents the averaged magnetization normalized by its maximal value, $m/m_{\text{max}}$, as a function of the filling $n$ and the interaction coupling strength $J$. The average magnetization is defined as $m=\sum_j\left\langle m_j\right\rangle / N$, where $N=833$ is the total number of lattice sites. The normalization factor $m_{\text{max}}=2-\left|4n-2\right|$ represents the theoretical upper bound derived from the occupation constraints~\cite{Dornellas2025arXiv}. For small $J$, the system settles into a trivial metallic state, as the local altermagnetization $\langle m_j \rangle$ vanishes during the self-consistent calculation. For larger $J$, a stable altermagnetic phase emerges with a finite order parameter $m$, which approaches its theoretical maximum $m/m_{\text{max}}\rightarrow 1$ in the strong-coupling regime.

\begin{figure}[t!]
\centering
\includegraphics[width=\columnwidth]{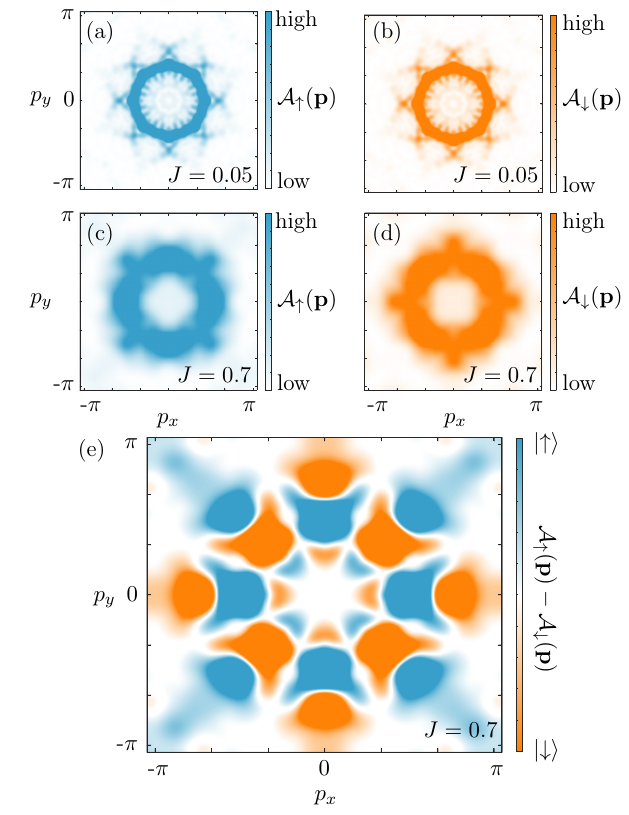}
\caption{Spectral density $\mathcal{A}(\bm{p})$ as a function of momentum $\bm{p}$. (a) and (c) show the spin-up spectral density $\mathcal{A}_{\uparrow}(\bm{p})$. (b) and (d) show the spin-down spectral density $\mathcal{A}_{\downarrow}(\bm{p})$. (e) presents the spin-resolved difference $\mathcal{A}_{\uparrow}(\bm{p}) - \mathcal{A}_{\downarrow}(\bm{p})$. We take the interaction
coupling $J = 0.05$ in (a) and (b), and $J = 0.7$ in (c)-(e). The electron filling is fixed at $n = 0.9$ in all panels.}
\label{fig_AB_spectrum}
\end{figure}

\begin{figure}[t!]
	\centering
	\includegraphics[width=\columnwidth]{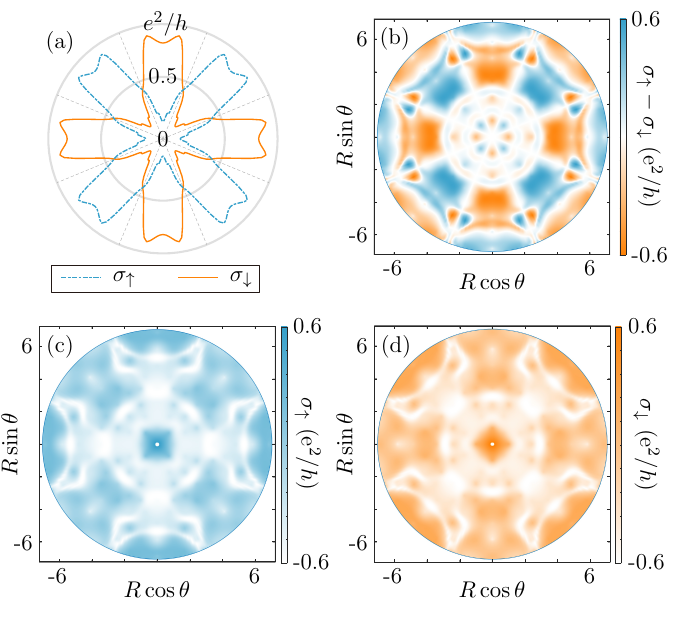}
	\caption{(a) Spin conductances $\sigma_{\uparrow}$ and $\sigma_{\downarrow}$ as functions of the angle $\phi$ with $R=4$. (b) $\sigma_{\uparrow}-\sigma_{\downarrow}$ as functions of the angle $\phi$ and the radius $R$. (c) and (d) depict the $\sigma_{\uparrow}$ and $\sigma_{\downarrow}$ as functions of the angle $\phi$ and the radius $R$. We take $J = 0.7$ and $n = 0.9$ in all panels.}
	\label{fig_AB_transport}
\end{figure}

\begin{figure*}[t!]
	\centering
	\includegraphics[width=2\columnwidth]{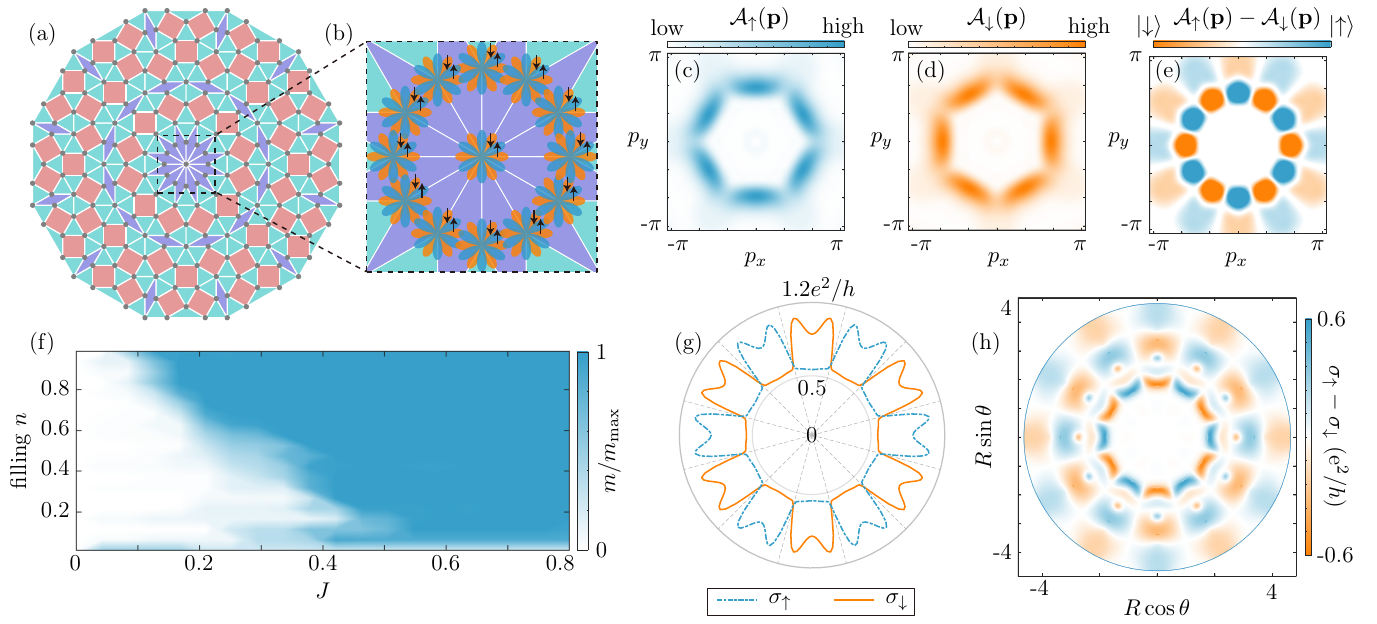}
	\caption{(a) Schematic illustration of the Stampfli-tiling dodecagonal quasicrystals. (b) Each site of the Stampfli-tiling quasicrystal features two orbitals connected by a $C_{12}$ rotation.  (c), (d), and (e)  describe the $\mathcal{A}_{\uparrow}(\bm{p})$, $\mathcal{A}_{\downarrow}(\bm{p})$, and $\mathcal{A}_{\uparrow}(\bm{p})-\mathcal{A}_{\downarrow}(\bm{p})$, respectively. (f) The normalized mean altermagnetization $m/m_{\max}$ as a function of the filling $n$ and the interaction coupling strength $J$. (g) Spin conductances $\sigma_{\uparrow}$ and $\sigma_{\downarrow}$ as functions of the angle $\phi$ with $R=4$. (h) $\sigma_{\uparrow}-\sigma_{\downarrow}$ as functions of the angle $\phi$ and the radius $R$. The electron filling is fixed at $n = 0.9$ in (c)-(e) and (g)-(h).}
	\label{fig_12}
\end{figure*}

{\emph{Spectral Function}.}---Direct visualization of spin-resolved momentum-space textures is experimentally challenging but has become increasingly feasible, even for complex materials like quasicrystals, thanks to the development of high-efficiency, multichannel spin detectors~\cite{Lv2019NRP,Sobota2021RMP,Wang2024MST}. Although quasicrystals lack discrete translational symmetry, meaning crystal momentum is not a good quantum number, their momentum-space properties can still be probed via Fourier analysis. Specifically, one can project eigenstates onto plane-wave-like trial states of the form:
$| \bm{p} s \mu\rangle= N \sum_{ \bm{x} _j} e^{i \bm{p} \cdot \bm{x} _j}\left| \bm{x} _j s \mu\right\rangle$.
These trial states serve as Fourier components of the real-space basis, representing spatial modulations with wave vector $\bm{p}$.
This Fourier-based projection thus provides a practical framework for analyzing the momentum-space structure of quasicrystalline systems in the absence of a Brillouin zone~\cite{Lesser2022PRR,Varjas2019}. We therefore compute the spin-resolved spectral function~\cite{Dornellas2025arXiv}
\begin{equation}
\mathcal{A} _s(\bm{p} )=-\frac{\delta}{\pi} \sum_{\lambda \mu} \frac{|\langle\lambda \mid \bm{p} s \mu\rangle|^2}{\left(\varepsilon_F-\varepsilon_\lambda\right)^2+\delta^2},
\end{equation}
where $\delta=10^{-2}$ is a small broadening factor and $\epsilon_F$ is the Fermi level corresponding to the filling number $n$. This quantity is directly accessible in spin-resolved angle-resolved photoemission spectroscopy (ARPES) measurements on quasicrystalline samples~\cite{Ahn2018Science,Rogalev2015NC}.

Figures~\ref{fig_AB_spectrum}(a)-\ref{fig_AB_spectrum}(b) show the numerically calculated spectra with $J=0.05$ [marked by the green star in Fig.~\ref{fig_AB_phase}(d)]. For this weak coupling, the system remains a trivial metal, and the spin-up and spin-down spectral functions are degenerate. The spectrum exhibits an eightfold rotational symmetry, reflecting the underlying symmetry of the quasicrystal lattice. The spectrum also displays fine hierarchical structures characteristic of quasiperiodic electronic spectra~\cite{JagannathanRMP,Bandres2016PRX}. In the following, we focus on the symmetry of the spin-resolved spectral weight, which provides the direct diagnostic of the altermagnetic phase.

Figures~\ref{fig_AB_spectrum}(c)-\ref{fig_AB_spectrum}(d) show the spectra for a strong interaction coupling of $J=0.7$ [marked by the yellow star in Fig.~\ref{fig_AB_phase}(d)]. In this regime, the system enters an altermagnetic phase, characterized by a finite self-consistent altermagnetic order. This results in a clear spin splitting in the spectral functions, accompanied by the breaking of $C_8$ symmetry while preserving a residual $C_4$ symmetry. Crucially, the spin-up and spin-down spectral functions are related by a $C_8$ rotation due to the preserved $C_8 T$ symmetry of the whole system. Consequently, the difference $\mathcal{A}{\uparrow}(\bm{p}) - \mathcal{A}{\downarrow}(\bm{p})$ transforms antisymmetrically under a $C_8$ rotation, meaning it changes sign and must vanish along the nodal line defined by $\theta = \pi/8 + n\pi/4$.

{\emph{Spin-Conductance}.}---Another distinguishing experimental feature of altermagnetism is its unique spin transport properties~\cite{sdongFeSb2,Nakaprb,shao2021spin,ifmmode2022Giant,Fernandestoplogical,zhang2024prl,Zhou2024PRL}.
We employ the two-terminal device to investigate the transport properties of the quasicrystalline system, where each lead is modeled as a tunable STM tip. One tip is positioned at $R(\cos\phi, \sin\phi)$, and the other is symmetrically located at $-R(\cos\phi, \sin\phi)$. By varying the probe separation $R$ and angle $\phi$, the setup allows angle-resolved probing of the transport anisotropy.

Specifically, each STM tip is modeled as a one-dimensional chain with the Hamiltonian
$
H_{\text{tip}} = \sum_{\langle jk \rangle} \Phi_j^{\dagger} t\, \tau_0 \sigma_0 \Phi_k,
$
where $\Phi_j$ denotes the spinor at site $j$ in the lead and $t=1$.
The coupling between the tip and the device is described by the tunneling Hamiltonian
$
H' = \sum_{jk} \Psi_j^{\dagger} t\, e^{-\bm{r}_{jk}} \tau_0 \sigma_0 \Phi_k+\text{h.c.},
$
where $\Psi_j$ denotes the spinor on the device [see Eq.~\eqref{Eq:Hamiltonian0}], and $\bm{r}_{jk}$ is the distance between the device site $j$ and the lead site $k$. The exponential factor ensures that the coupling is spatially localized, restricting the tunneling to sites in close proximity to the tip position. The conductance between the two tips is calculated by using the Landauer-B\"uttiker formula~\cite{Landauer1970Philosophical,Buttiker1988PRB,
Fisher1981PRB} and the recursive Green's
function method \cite{Mackinnon1985Zeitschrift,Metalidis2005PRB}.

Figure~\ref{fig_AB_transport}(a) shows the spin conductance $\sigma_{\uparrow,\downarrow}$ as functions of angle $\phi$ for $R=4$. The spin-dependent conductance exhibits strong anisotropy, a phenomenon more clearly visualized in Figs.~\ref{fig_AB_transport}(c)-(d), where we plot $\sigma_{\uparrow,\downarrow}$ as functions of $R\cos\theta$ and $R\sin\theta$.
Similar to the spectral function, the conductance for each spin component displays a clear $C_4$ symmetry. However, a $C_8$ rotation transforms the spin-up conductance into the spin-down conductance. As a result, the difference $\sigma_{\uparrow} - \sigma_{\downarrow}$ changes sign under an eight-fold rotation and must vanish along the nodal line $\theta = \pi/8 + n\pi/4$ [see Fig.~\ref{fig_AB_transport}(b)]. These transport signatures further confirm that the system realizes a $g$-wave altermagnet.
While experimentally demanding, multi-probe STM techniques provide a unique and powerful route to probe non-local transport anisotropies and correlation functions. Our calculations thus provide a clear theoretical target and a strong motivation for such advanced measurements, where the predicted symmetry of the spin-conductance offers an unambiguous signature of the underlying quasicrystalline altermagnetic order.

{\emph{Stampfli-tiling dodecagonal quasicrystal}.}---We extend our analysis to the Stampfli-tiling quasicrystal~[Fig.~\ref{fig_12}(a)], which features a global twelve-fold rotational symmetry~\cite{Ahn2018Science}. This tiling consists of three primitive tiles: squares (red), regular triangles (purple), and rhombuses (cyan) with an acute angle of $30^{\circ}$. The Hamiltonian is analogous to the AB-tiling case, but with $\eta=3$ in Eqs.~\eqref{Eq:Hamiltonian0}-\eqref{Eq:hopping} to reflect the $C_{12}$ symmetry. This can be realized by choosing the $f_{x\left(x^2-3 y^2\right)}$ and $f_{y\left(3 x^2-y^2\right)}$ orbitals. Similarly, these two $f$-orbitals form $E_3$ doublets in dodecagonal ($C_{12v}$) quasicrystals~\cite{Socolar1989PRB,Madison2013PSS}. In this setup, for $\theta=0$, hopping is dominated by the overlap $t_1$ between the first orbitals and the second orbital have a smaller overlap $t_2$ [Figs.~\ref{fig_12}(a)-\ref{fig_12}(b)]. For $\theta=\pi/6$, the roles are reversed.

Figure~\ref{fig_12}(f) shows the normalized altermagnetization $m/m_{\text{max}}$ as a function of filling $n$ and interaction coupling strength $J$. We find that the altermagnetic phase in the Stampfli-tiling quasicrystal remains stable at large $J$, with $m$ approaching its theoretical maximum. This behavior closely resembles our findings for the AB-tiling quasicrystal, indicating that the emergence of altermagnetism is a robust feature in quasicrystalline systems with higher-fold rotational symmetries.

Figures~\ref{fig_12}(c)-\ref{fig_12}(e) and \ref{fig_12}(g)-\ref{fig_12}(h) present the spectral functions and transport properties of the Stampfli-tiling system, respectively. Due to the underlying $C_{12}$ symmetry, the spin-up and spin-down components of the spectra and transport response are related by a twelve-fold rotation. As a result, their differences, i.e., $A_{\uparrow}(\bm{p}) - A_{\downarrow}(\bm{p})$ and $\sigma_{\uparrow} - \sigma_{\downarrow}$, exhibit antisymmetric behavior under $C_{12}$ rotations, changing sign upon a $\pi/6$ rotation. These signatures establish the presence of a $C_{12}T$ protected $i$-wave altermagnetic phase in this quasicrystal.

{\emph{Conclusion and Discussion}.}---We have extended the concept of altermagnetism to quasicrystals, demonstrating that their unique higher-fold rotational symmetries, incompatible with translational periodicity, can protect novel altermagnetic phases. By calculating symmetry-protected spectral textures and transport features, we have identified clear experimental signatures. The characteristic spectral textures are directly accessible via spin-resolved ARPES, and the anisotropic transport signatures can be probed using multi-tip STM.


We acknowledge that resolving these nodal structures via spin-ARPES or double-tip STM represents a significant challenge with current techniques applied to solid-state quasicrystals~\cite{Lv2019NRP,Sobota2021RMP,Wang2024MST,sdongFeSb2,Nakaprb,shao2021spin,ifmmode2022Giant,Fernandestoplogical,zhang2024prl,Zhou2024PRL}. However, the identification of candidate materials like Au-In-Eu provides a clear motivation for future spectroscopic efforts. Furthermore, the predicted phases may be simulated in highly tunable platforms, such as optical quasicrystals for ultracold atoms~\cite{Jagannathan2013EPL,Viebahn2019PRL} or moir\'{e} quasicrystals~\cite{Uri2023Nature,LiYX2024Nature}, where the specific orbital-dependent interactions can be engineered directly.

The proposed mechanism is experimentally motivated by two developments. First, transition-metal quasicrystals and approximants~\cite{Jiang1991PRL} provide quasiperiodic environments in which orbital doublets and direction-dependent hopping textures may arise. Second, the recent observation of long-range antiferromagnetic order in  Au$_{56}$In$_{28.5}$Eu$_{15.5}$~\cite{Tamura2025NatPhys} demonstrates that coherent magnetic order can be stabilized on quasiperiodic lattices. These results suggest that quasicrystalline magnets are promising systems in which to search for altermagnetic spin splitting, while a fully material-specific microscopic model remains a subject for future work.



\begin{acknowledgments}
This work was supported by the NSFC (under Grants Nos.~12474151, 92565103, 12304195, 12547101, and U25D8012). D.H.X. acknowledges the Beijing National Laboratory for Condensed Matter Physics (No. 2024BNLCMPKF025) and the Fundamental Research Funds for the Central Universities (Grant No. 2025CDJIAISYB-032), the Natural Science Foundation of Chongqing (Grant No. CSTB2025NSCQ-LZX0010). R.C. acknowledges the Chutian Scholars Program in Hubei Province, the Hubei Provincial Natural Science Foundation (Grant No. 2025AFA081), the key project of Hubei provincial department of education (under Grant No. D20241004), and the original seed program of Hubei University. B.Z. acknowledges the Wuhan city key R\&D program (under Grant No. 2025050602030069).
\end{acknowledgments}

%
%
%
\bibliographystyle{apsrev4-1-etal-title_6authors}
\bibliography{refs-transport,refs-transport_v1}

\section*{End Matter}
The $g$-wave and $i$-wave altermagnetic phases found here can be understood within the magnetic, or black-and-white group classification of quasiperiodic crystals developed by R. Lifshitz~\cite{spingroupinQCs,Lifshitz2024ECMP,Lifshitz1997RMP}. This framework classifies magnetic structures by combining spatial operations with time reversal or spin reversal. In the present context, it provides a symmetry-based basis for analyzing which magnetic symmetries of a quasiperiodic crystal can allow altermagnetic spin splitting in electronic spectra.

In Sec.~SII D of the Supplemental Material~\cite{Supp}, we adapt the spin-group language used for periodic altermagnets~\cite{spingroupCS1,spingroupCS2,spingroupCS3,Smejkal22PRX,ZengSK2024PRB} to two-dimensional quasiperiodic point groups. We identify the admissible altermagnetic phases through the decomposition of the real-space symmetry group $\mathcal{G}$:
\begin{equation}
R_3 = [\mathcal{E} || \mathcal{H}] + [\mathcal{C}_2 || \mathcal{G} - \mathcal{H}].
\end{equation}
Here, the spin group $R_3$ is constructed by pairing spatial operations in the halving subgroup $\mathcal{H}$ with the spin identity $\mathcal{E}$, while operations in the coset $\mathcal{G}-\mathcal{H}$ are coupled with a spin flip $\mathcal{C}_2$.

Based on this formalism, we categorize quasicrystals into three classes distinguished by the order of their rotational symmetry, odd-, $4n$-, and $(4n+2)$-fold. Specifically, we show that while mirror-driven altermagnetism is universally accessible, pure rotation-driven altermagnetism is allowed for $4n$-fold systems (like the octagonal case studied here). This distinction arises because only $4n$-fold symmetries allow the vertical $180^{\circ}$ rotation ($C_{2z}$) to be sequestered within the spin-preserving subgroup $\mathcal{H}$, thereby avoiding the enforced spin degeneracy that forbids rotation-driven phases in odd and $(4n+2)$-fold systems.  The Ammann-Beenker and Stampfli tilings used in this work possess the full dihedral point symmetries $8mm$ and $12mm$, respectively. Accordingly, the magnetic point groups of the altermagnetic phases realized in these tilings are $8^{\prime}m^{\prime}m$ and $12^{\prime}m^{\prime}m$. In the cases studied here, the twofold rotation can be placed in the spin-preserving subgroup, allowing altermagnetic spin splitting protected by $C_{8}T$ and $C_{12}T$.
\end{document}